# Amorphous Silicon Solar cells with a Core-Shell Nanograting Structure

*L. Yang[1], L. Mo[1], Y. Okuno[2], and S. He[1]*

[1]COER, Zhejiang University, China

[2]Kumamoto University, Japan

**ABSTRACT**

We systematically investigate the optical behaviors of an amorphous silicon solar cell based on a core-shell nanograting structure. The horizontally propagating Bloch waves and Surface Plasmon Polariton (SPP) waves lead to significant absorption enhancements and consequently short-circuit current enhancements of this structure, compared with the conventional planar one. The perpendicular carrier collection makes this structure optically thick and electronically thin. An optimal design is achieved through full-field numerical simulation, and physical explanation is given. Our numerical results show that this configuration has ultrabroadband, omnidirectional and polarization-insensitive responses, and has a great potential in photovoltaics.

## 1. INTRODUCTION

Photovoltaics, converting sunlight to electricity, has become a most promising technology for green and renewable energy. Traditional crystal silicon (c-Si)-based solar cells, with ~ 80% market occupation, have good power conversion efficiency (~ 25% [1]), but the cost is very high (about 40% of the cost is from the c-Si material itself). Since c-Si is an indirect-bandgap semiconductor, the film has to be thick (usually more than 200 μm) to absorb as much solar energy as possible. In addition, to achieve a high photocarrier collection efficiency, expensive multiple purification processes are required to obtain a solar-grade c-Si. In contrast, low-grade hydrogenated amorphous silicon (a-Si:H) with long range disorder in its atomic structure, behaves like a direct-bandgap semiconductor with two-order higher optical absorption coefficient as compared with its crystalline counterpart. Only 1 μm thick a-Si:H layer can absorb more than 90% sunlight [2, 3]. However, too many dangling bonds inside the material make the carrier diffusion length (~ 300 nm) much shorter than its light absorption length. Moreover, a thinner amorphous film (below 300 nm) is preferred to reduce the Staebler-Wronski light-induced degradation effect [4]. Due to such a large mismatch in length, the power conversion efficiency can not be as high as what can be achieved with the c-Si solar cells [1]. In this case, light trapping becomes critically important to overcome this tradeoff and finally improve the whole device performance both optically and electronically.

Orthogonalizing the sunlight absorption direction and the photocarrier collection direction is a most promising method to make optically thick and electronically thin solar absorber. This can directly overcome the length tradeoff faced by most thin-film solar cells. It is possible to convert the normally incident sunlight into SPP waves propagating horizontally along the metal-dielectric interface. Only the near-field profiles in the semiconductor contribute to the generation of photocarriers. In this case, the solar absorber can be much thinner to guarantee the complete collection of photocarriers [5-7]. Some a-Si:H solar cells based on nanocoax [8], nanodome [9, 10], grating coupler [11] and nanostructured back reflectors [12] have been reported with various shapes of active layers to enhance the short-circuit current. An alternative scheme is to use Si micro-/nano-wire [13-17] or nano-hole arrays [18, 19] with axial light absorption and radial carrier collection. As long as the arrays are long enough in the axial direction, the sunlight could be totally absorbed without affecting the carrier collection efficiency in the radial direction. Therefore, such micro-/nano-wires or nano-hole arrays are considered as one of the most potential platforms for the next-generation photovoltaics [20]. Aiming to combine the advantages of both schemes, we investigate in this paper an a-Si:H solar cell based on core-shell nanograting. Our cell has a semiconducting core as the absorbing material, and a metallic shell as both an electrode and an essential plasmonic element. The basic core-shell structure used in photovoltaic applications was first proposed by Zhu *et al.*, who demonstrated this concept using an a-Si:H Schottky core-shell nanowire solar cell [21]. More than 50% increase of the short-circuit current (as compared with the conventional planar thin-film counterpart) was achieved due to the efficient carrier collection capability. Only at the end of their paper did Zhu et al. mention briefly that the n-i-p based core-shell nanowires should have better electronic performances than the Schottky junction itself. There were no optical/physical interpretations. Here in this paper, we will give a systematic numerical study on this core-shell nanograting based solar cell from the photonic aspect and optimize the design to obtain a high short-circuit current enhancement. Our simulation results show that this design has ultra-broadband, omnidirectional and polarization-insensitive absorption responses with much better carrier collection efficiency. Therefore, we could expect wide potential applications of this design in photovoltaics.

**2. Structure and Numerical Simulation Method**

The schematic of our core-shell nanograting based a-Si:H solar cell is shown in Figure 1a. In this structure, the dielectric nanoribs including the absorbing a-Si:H and the transparent ZnO materials are seen as the cores and the Au film covering them is the shell. The sunlight going through the glass substrate will be tightly trapped in the dielectric core and absorbed in the absorbing a-Si:H layers, generating useful photocarriers. For comparison, a conventional planar a-Si:H based solar cell is plotted in Figure 1b. For fair comparisons, the a-Si:H thickness of the planar structure is set to $h'_{Si}$, which is not equal to $h_{Si}$, so that $h'_{Si}$ can be tuned to ensure that both structures in comparison have the same absorbing area.

We perform quantitative full-field electromagnetic simulations with the Finite-Difference Time Domain (FDTD) method by using the commercial software of Lumerical FDTD solutions. For a solar cell, the light absorption in the a-Si:H layers, $\eta(\lambda)$ expressed in Eq. (1) below, is an important parameter, which determines the short-circuit current density of the whole device. Here, we assume that the photocarriers will be completely collected by the electrodes. Then the short-circuit current density, $J_{sc}$, can easily be calculated by

$$\eta(\lambda) = \frac{\iint_{\text{a-Si:H}} \frac{\pi c}{\lambda} \cdot \mathrm{Im}(\varepsilon) \left| E \right|^2 dxdy}{\text{Source Power}} \tag{1}$$

$$J_{sc} = e \int_{400\text{ nm}}^{900\text{ nm}} \frac{\lambda}{hc} \cdot \eta(\lambda) \cdot I_{AM1.5}(\lambda) d\lambda \tag{2}$$

where $\mathrm{Im}(\varepsilon)$ is the imaginary part of the dielectric constant of a-Si:H, $c$ is the light speed in vacuum, $h$ is the Planck's constant, and $I_{AM1.5}$ is the AM1.5 solar spectrum.

**3. Simulation Results**

We first calculated the absorption spectra of both TE- and TM-polarizations for our solar cell design in Figure 1a with $w_r$ = 100 nm, $h_r$ = 60 nm, $h_{sl}$ = 80 nm, $h_{Si}$ = $w_{Si}$ = 50 nm, and $P$ = 400 nm and plotted them in Figure 2a. For comparison, we also plotted the absorption spectrum for the conventional planar solar cell schematically in Figure 1b, and the single-pass absorption spectrum in a semi-infinite a-Si:H structure calculated with the same monitor area as the a-Si:H region in Figure 1b. Figure 2b shows the former three absorption spectra normalized to the single-pass absorption in order to eliminate the a-Si:H's intrinsic absorption and to give a clear view of the structure-induced absorption enhancement. Compared with the single-pass absorption in Figure 3a, the conventional planar solar cell with back Au reflector has higher absorption throughout the whole wavelength range due to the strong reflection. However, it is still lower than that of our solar cell design for both TE- and TM-polarizations. Our design shows even better performances in the longer wavelength range from 550 nm to 900 nm. Such a broad band (over 350 nm) of absorption enhancement is much wider than the previously reported values of e.g., ~ 250 nm for the solar cell with back metallic nanograting [7] and ~ 300 nm for nano-coax and nanowire solar cells [8, 13].

When impinging to the core-shell nanograting based solar cell normally from the bottom as shown in Figure 1a, the light scattered by the nanograting obtains a horizontal wave vector, $k_x = m \cdot 2\pi/P$ (m is an integer), which enables the generation of photonic waves via phase-matching conditions. At the same time, the propagating waves inside the structure are apt to be confined in the high-index dielectric layers in the $y$ direction rather than the low-index glass substrate. For TE polarization, three Bloch modes are clearly observed at absorption peak wavelengths, namely, $\lambda$ = 588, 681, and 858 nm (Figure 3a), whose electric field $|E_z|$ distributions are shown in Figures 3a, b and c, respectively. For the mode at $\lambda$ = 858 nm, $|E_z|$ is very strong in the a-Si:H layer compared with the modes at $\lambda$ = 588 and 681 nm, leading to the strongest enhancement of light absorption in Figure 2. For TM polarization, the situation is different. In Figures 3d, e, f, and g we plotted the magnetic field $|H_z|$ distributions at the absorption peak wavelengths of 585, 711, 795 and 872 nm (Figure 2a), respectively. There are two Bloch modes at shorter wavelengths of 585 and 711 nm. At longer wavelengths, SPP waves are fully excited to propagate along the interface between the Au and a-Si:H layers. Due to the constructive interferences between SPPs propagating along $+x$ and $-x$ directions, two resonance-induced absorption peaks are observed in Figure 2a, corresponding to two different optical field distributions shown in Figures 3f and g. These tightly-confined optical fields are much stronger than the short-wavelength Bloch modes shown in Figures 3d and e, resulting in much higher absorptions of light shown in Figure 2b. Though the Bloch mode at $\lambda$ = 711 nm is weakly coupled to the SPP waves, the absorption is still much higher than the pure Bloch mode at $\lambda$ = 585 nm. Because of these modes, the absorption spectra are enhanced and broadened significantly for both TE- and TM-polarizations. Since SPP waves have higher losses due to the Joule heat in the metal film, the absorption density for TM-polarization is slightly lower than those for TE-polarization as seen in Figures 2.

Below we study the structural effects. The short-circuit current density will be further improved by tuning the structural parameters. If the nanograting period varies, the horizontal wave vector, $k_x$, will be changed and, accordingly, the resonance-induced absorption band will be shifted based on the phase-matching condition. If the absorption band matches well the AM 1.5 solar spectrum, the short-circuit current density should become high. It is clearly seen from Figure 4a that at period $P$ = 380 nm the enhancement of short-circuit current density reaches peak values $J_{sc_enh}$ = 55.11%, 43.09% and 47.1% for TE-, TM-, and hybrid polarizations, respectively (for the case of $w_{Si}$ = 50 nm). Below we give some explanation for some structural effects based on our calculated absorption spectra (not shown here to avoid too many figures). Compared with the case of $P$ = 400 nm in Figure 2, the rightmost absorption peak at the longest wavelength fully moves into the main solar spectral range and thus improves the short-circuit current density more for both TE- and TM-polarizations. As the nanograting period decreases further, the absorption spectra blue shift further. It is interesting to see that at $P$ = 250 nm there is a local maximum of $J_{sc_enh}$ for TM-polarization. At this point, the leftmost absorption peak in the short wavelength range moves out of the solar spectral range, while those in the long wavelength range blue shift to cover the solar spectrum maximum. The short-circuit current enhancement is moderately high for TM-polarization, but there is no local maximum for TE-polarization in the short period range.

For the conventional planar solar cell shown in Figure 1b, the TCO layer works both as a transparent electrode and as an anti-reflection film. Similarly, for our own design shown in Figure 1a, the TCO slab has the same functions. Therefore, there should be an optimal thickness for this layer to trap more light in the absorbing material. Here we performed an optimization procedure for both solar cell designs and showed the results in Figure 5. Other structural parameters are chosen to the optimized values, i.e., $h_{Si}$ = 50 nm, $w_r$ = 100 nm, $h_r$ = 60 nm, $w_{Si}$ = 40 nm and $P$ = 380 nm. From this figure, we see that the conventional planar solar cell has a maximal $J_{sc}$ = 143.98 mA/cm$^2$ at $h_{sl}$ = 62 nm, while our design has maximal values of $J_{sc}$ = 209.16 mA/cm$^2$ at $h_{sl}$ = 70 nm for hybrid polarization. Therefore, our design has much better performance than the planar design (even with the best antireflection).

To this end, we have obtained all the optimal structural parameters for our solar cell design: $h_{Si}$ = 50

nm, $w_r$ = 100 nm, $h_r$ = 60 nm, $P$ = 380 nm, $w_{Si}$ = 40 nm, and $h_{sl}$ = 70 nm. For this optimal design, we evaluated the dependence of incident angle $\theta$ defined in Figure 1, by calculating short-circuit current density enhancement $J_{sc_enh}$ with respect to the conventional planar design. The result is shown in Figure 6, from which one sees that the short-circuit current density is enhanced by 10% - 60% over a broad incident angle (even at $\theta$ which is as large as ±80°) for both TE- and TM-polarizations. Such remarkable wide-angle absorption enhancement of our core-shell nanograting based solar cell is comparable to that of the structure with back metallic nanograting [7] and much better than that of the structure with back metallic nanogrooves [5]. Therefore, our design can be considered as an omnidirectional solar absorber, which is much more favorable for practical implementations than the planar design. When $\theta$ is greater than ±80°, $J_{sc_enh}$ becomes larger for TM- and hybrid polarizations, but $J_{sc}$ is too low for both solar cell designs (not shown here). In Figure 7, $J_{sc}$ is enhanced over a large incident angle range regardless of the sunlight polarizations. In this respect, our solar cell with core-shell nanograting is polarization independent though it is based on a 1-D nanograting.

## 4. Conclusion

In conclusion, we have numerically investigated the optical behaviors of the core-shell nanograting based a-Si:H solar cell and given an optimal design. The simulation results show that this design compared with the conventional one has much better performances of absorption and carrier generation in terms of ultrabroadband, omnidirectional, polarization-independent responses. The core-shell structure is also favorable for carrier collection [37]. Therefore, we could expect wide practical implementations of this configuration.

**Figures**

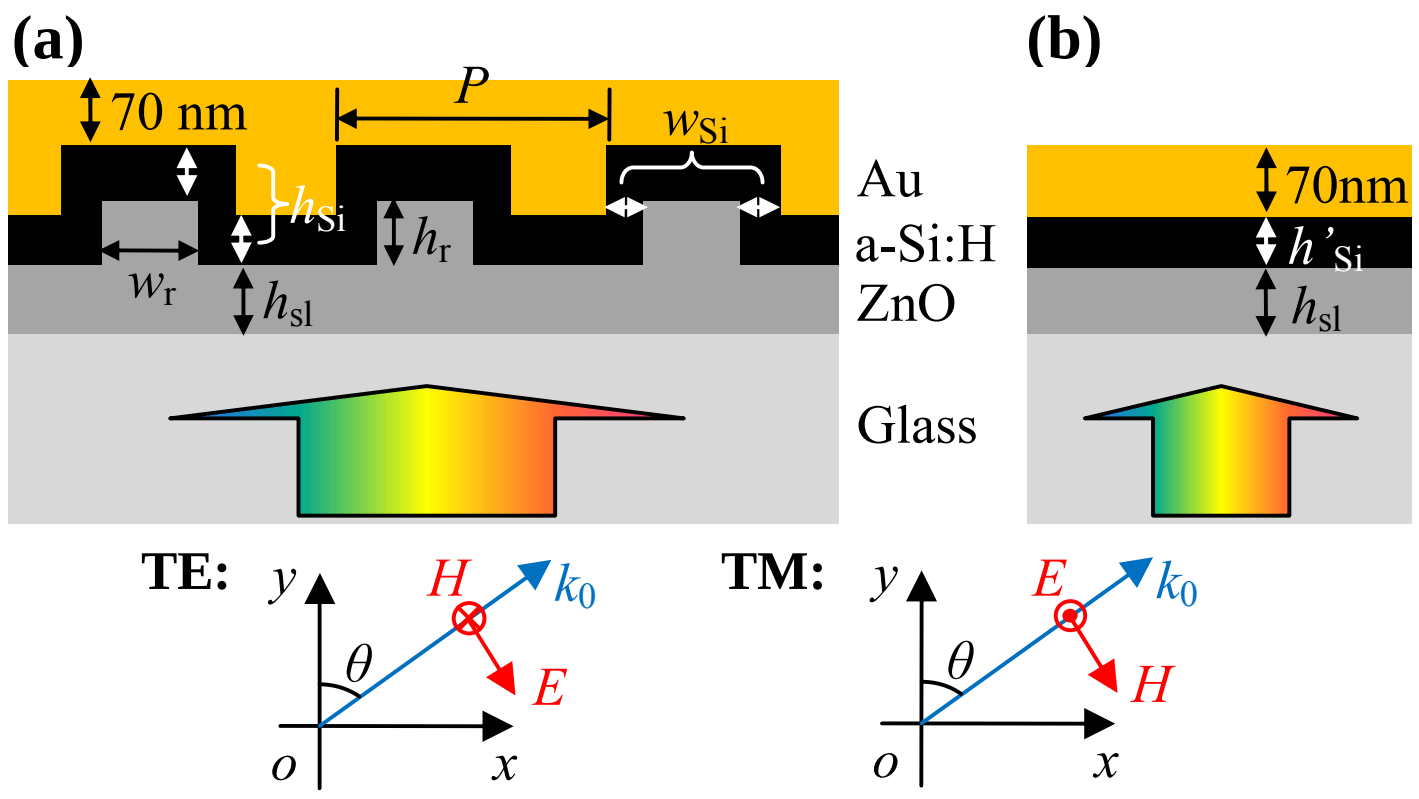

**Figure 1.** The schematic diagrams of (a) an original a-Si:H solar cell based on core-shell nanograting; and (b) a conventional planar a-Si:H solar cell used for comparison. The sunlight is incident from the glass substrate with an incident angle, $\theta$, and a wave vector, $k_0$. The directions of the electric field ($E$) and magnetic field ($H$) are indicated at the bottom for both TE- (Transverse Electric) and TM- (Transverse Magnetic) polarizations.

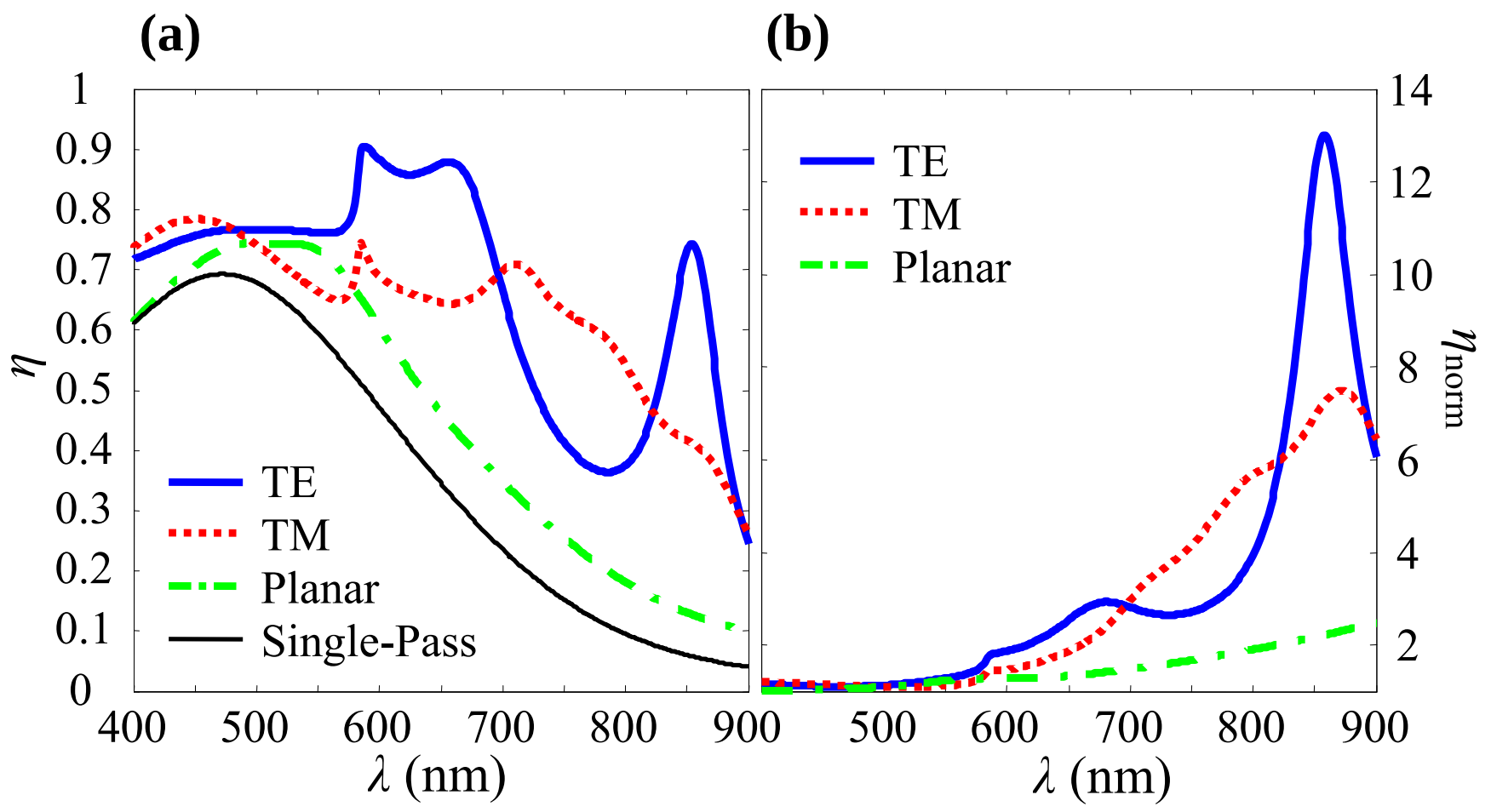

**Figure 2.** (a) Absorption spectra $\eta$ of our structure, including both TE- (blue thick solid line) and TM- (red dotted line) polarizations, and the conventional planar structure (green dash-dotted line) and single-pass absorption spectrum in a semi-infinite a-Si:H structure (black thin solid line); (b) Normalized absorption spectra, $\eta_{norm}$, with respect to the single-pass absorption. The structural parameters are $w_r$ =

100 nm, $h_r$ = 60 nm, $h_{sl}$ = 80 nm, $h_{Si}$ = $w_{Si}$ = 50 nm, and $P$ = 400 nm.

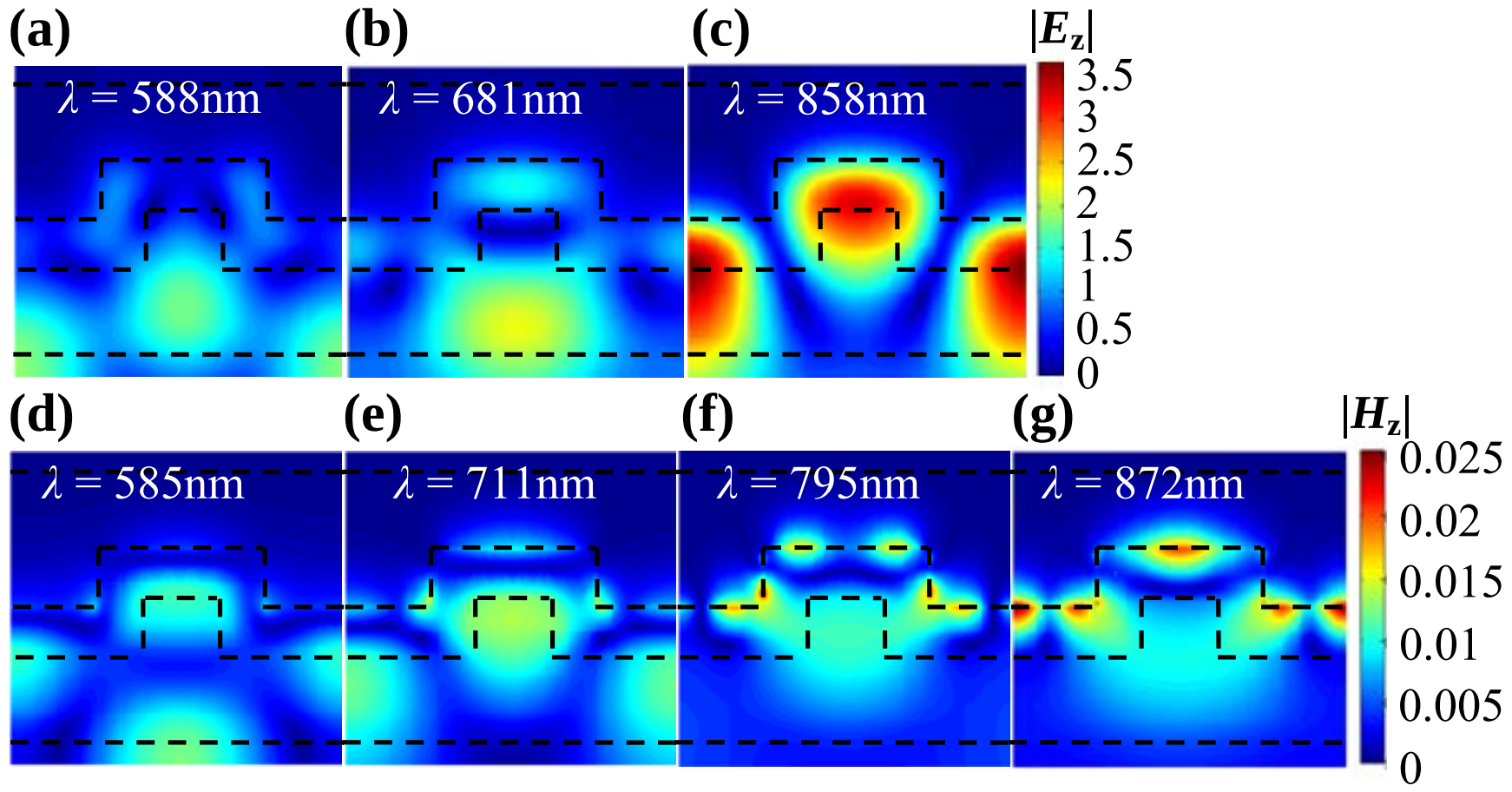

**Figure 3**. Distributions of electric field $|E_z|$ for TE-polarization at absorption peak wavelengths: (a) $\lambda$ = 588 nm, (b) $\lambda$ = 681 nm, and (c) $\lambda$ = 858 nm; and magnetic field $|H_z|$ for TM-polarization at absorption peak wavelengths: (d) $\lambda$ = 585 nm, (e) $\lambda$ = 711 nm, (f) $\lambda$ = 795 nm, and (g) $\lambda$ = 872 nm. The dashed lines in these figures denote the core-shell nanograting in one period. The structural parameters are $w_r$ = 100 nm, $h_r$ = 60 nm, $h_{sl}$ = 80 nm, $h_{Si}$ = $w_{Si}$ = 50 nm, and $P$ = 400 nm.

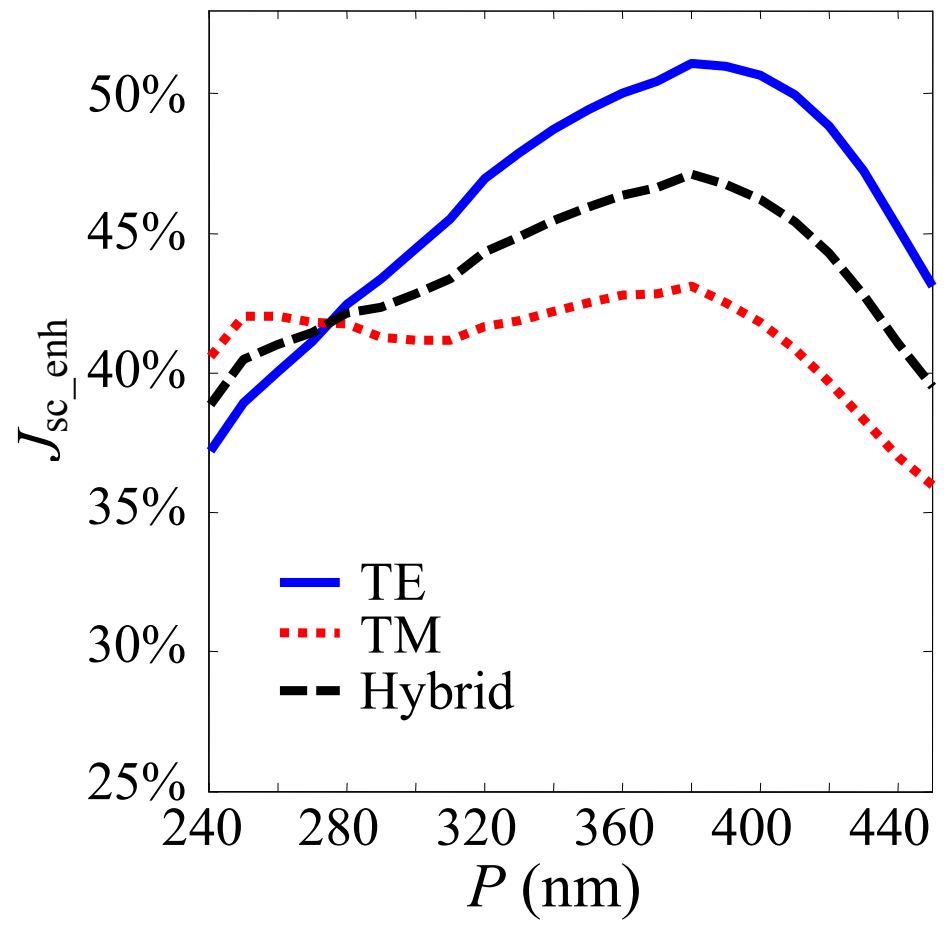

**Figure 4.** Short-circuit current density enhancement, $J_{sc_enh}$, of our design as a function of the nanograting period, $P$, for TE- (blue solid line), TM- (red dotted line) and hybrid (black dashed line)

polarizations. The other structural parameters are $w_{Si}$ = 50 nm, $h_{Si}$ = 50 nm, $w_r$ = 100 nm, $h_r$ = 60 nm, $h_{sl}$ = 80 nm.

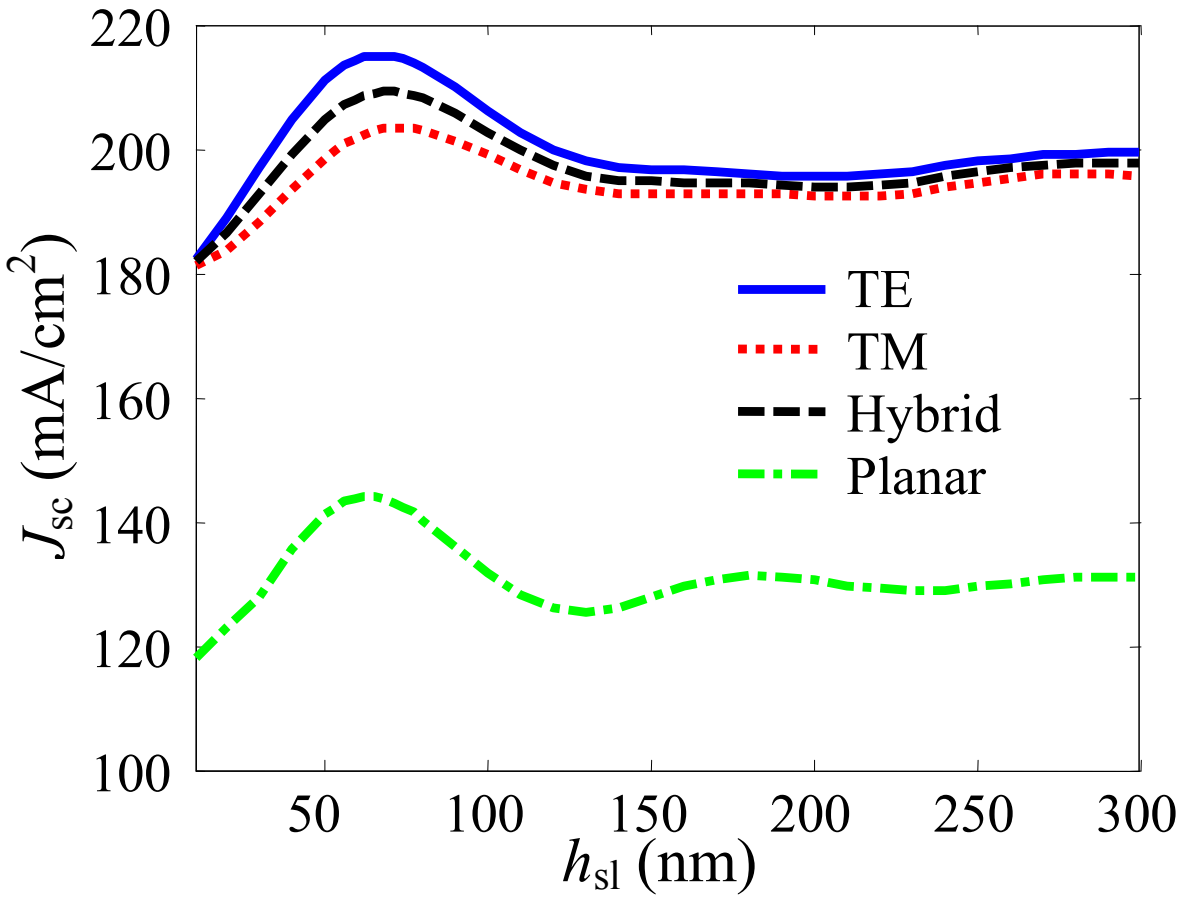

**Figure 5.** Short-circuit current density, $J_{sc}$, as a function of the TCO slab thickness, $h_{sl}$, for the solar cell with core-shell nanograting for TE- (blue solid line), TM- (red dotted line) and hybrid (black dashed line) polarizations and for the conventional planar solar cell (green dash-dotted line). The other structural parameters are $h_{Si}$ = 50 nm, $w_r$ = 100 nm, $h_r$ = 60 nm, $w_{Si}$ = 40 nm and $P$ = 380 nm.

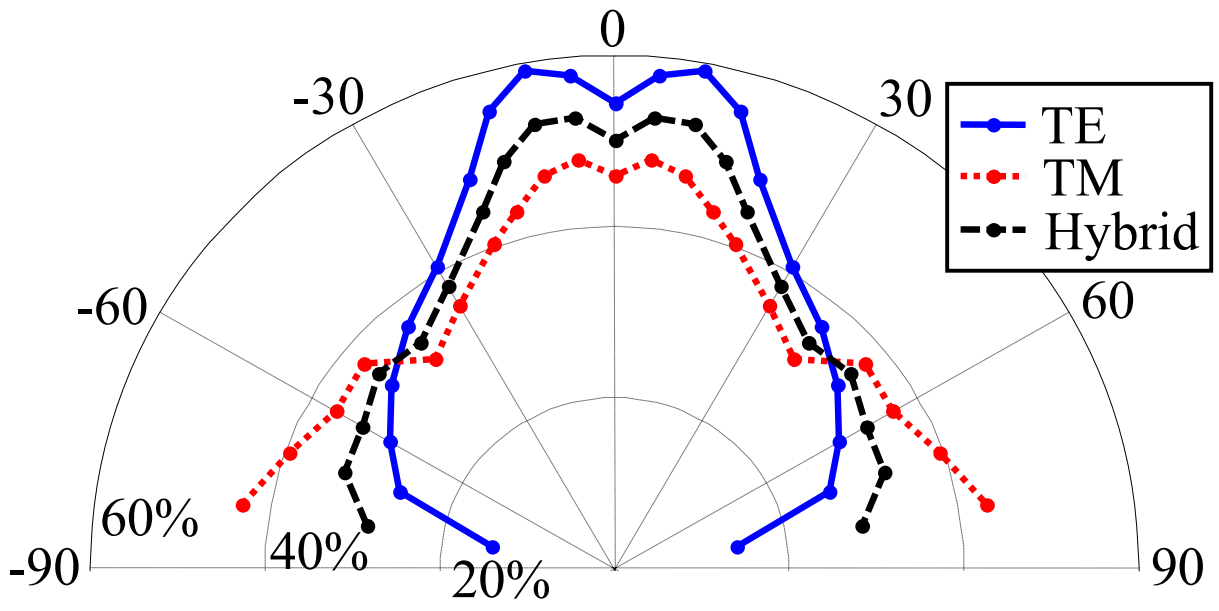

**Figure 6.** The dependence of short-circuit current density enhancement, $J_{sc_enh}$, on the incident angle, $\theta$, for TE- (blue solid line), TM- (red dotted line) and hybrid (black dashed line) polarizations. ($J_{sc_enh}$ is calculated with respect to the conventional planar solar cell.) The structural parameters are $h_{Si}$ = 50 nm, $w_r$ = 100 nm, $h_r$ = 60 nm, $P$ = 380 nm, $w_{Si}$ = 40 nm, and $h_{sl}$ = 70 nm.